\documentclass[12pt]{iopart}

\usepackage{graphicx}
\usepackage{dcolumn}
\usepackage{bm}

\begin{document}






\def\beq{\begin{equation}}
\def\eeq{\end{equation}}
\def\bea{\begin{eqnarray}}
\def\eea{\end{eqnarray}}
\def\ben{\begin{enumerate}}
\def\een{\end{enumerate}}
\def\la{\langle}
\def\ra{\rangle}
\def\a{\alpha}
\def\b{\beta}
\def\g{\gamma}\def\G{\Gamma}
\def\d{\delta}
\def\e{\epsilon}
\def\phi{\varphi}
\def\k{\kappa}
\def\l{\lambda}
\def\m{\mu}
\def\n{\nu}
\def\o{\omega}
\def\p{\pi}
\def\r{\rho}
\def\s{\sigma}
\def\t{\tau}
\def\L{{\cal L}}
\def\S{\Sigma }
\def\gsim{\; \raisebox{-.8ex}{$\stackrel{\textstyle >}{\sim}$}\;}
\def\lsim{\; \raisebox{-.8ex}{$\stackrel{\textstyle <}{\sim}$}\;}
\def\gtrsim{\gsim}
\def\lessim{\lsim}
\def\loc{{\rm local}}
\def\vm{v_{\rm max}}
\def\bh{\bar{h}}
\def\del{\partial}
\def\nab{\nabla}
\def\half{{\textstyle{\frac{1}{2}}}}
\def\fourth{{\textstyle{\frac{1}{4}}}}

\def\static{\cite{Eling:2006df}}

\title{Black Holes in Einstein-Aether Theory}

\author{Christopher Eling and Ted Jacobson}
\address{Department of Physics, University of Maryland\\ College Park, MD 20742-4111 USA}
\ead{cteling@physics.umd.edu}
\ead{jacobson@umd.edu}


\begin{abstract}

We study black hole solutions in general relativity coupled to a
unit timelike vector field dubbed the ``aether". To be causally
isolated a black hole interior must trap matter fields as well as
all aether and metric modes. The theory possesses spin-0, spin-1,
and spin-2 modes whose speeds depend on four coupling coefficients.
We find that the full three-parameter family of local spherically
symmetric static solutions is always regular at a metric horizon,
but only a two-parameter subset is regular at a spin-0 horizon.
Asymptotic flatness imposes another condition, leaving a
one-parameter family of regular black holes. These solutions are
compared to the Schwarzschild solution using numerical integration
for a special class of coupling coefficients. They are very close to
Schwarzschild outside the horizon for a wide range of couplings, and
have a spacelike singularity inside, but differ inside
quantitatively. Some quantities constructed from the metric and
aether oscillate in the interior as the singularity is approached.
The aether is at rest at spatial infinity and flows into the black
hole, but differs significantly from the the 4-velocity of
freely-falling geodesics.

\end{abstract}

\maketitle

\section{Introduction}
\label{intro}

Einstein-Aether theory,  nicknamed ``ae-theory", consists of general
relativity (GR) coupled to a dynamical, unit timelike vector field
$u^a$ called the ``aether". The aether is somewhat like a nonlinear
sigma model field with a hyperbolic target space. But, unlike for a
sigma model, the hyperboloid is dynamically determined at each point
by the spacetime metric and the aether couples to the metric via
covariant derivatives. The aether dynamics thus modifies the metric
dynamics, and the coupled system differs significantly from the
familiar examples of GR coupled to matter fields.

Properties of ae-theory have been extensively studied for the past
few years, and many observational signatures have been worked out
(see \cite{Eling:2004dk} for a recent review). The primary
motivation for studying ae-theory is to explore the gravitational
consequences of local Lorentz symmetry violation. The aether itself
breaks local boost invariance at each point, while preserving
rotational symmetry in a preferred frame. There are four free
coupling coefficients in the general ae-theory Lagrangian (besides
the gravitational constant), thus providing a somewhat general class
of boost violating models\footnote{For a much broader class of
models with Lorentz violating gravity see
\cite{Kostelecky:2003fs,Bailey:2006fd}.}. There is a common range of
the four coupling coefficients for which simultaneously all
post-Newtonian parameters agree with those of GR~\cite{Eling:2003rd,
Graesser:2005bg,Foster:2005dk}, where all perturbations are
stable~\cite{Jacobson:2004ts,Lim:2004su} and have positive
energy~\cite{Lim:2004su,Eling:2005zq,Foster:2006az}, gravitational
and aether vacuum Cerenkov radiation~\cite{Elliott:2005va} is
absent~\cite{Foster:2005dk}, the gravitational constant in the
Friedman equation agrees with Newton's
constant~\cite{Carroll:2004ai}, and gravitational radiation damping
of binary pulsar orbits agrees with the rate in GR for weak fields
at lowest post-Newtonian order~\cite{Foster:2006az}.

In spherical symmetry the aether has one degree of
freedom, a radial tilting. Restricting to the time-independent
case, we found in \static~that there is a three-parameter family
of local solutions, of which a two-parameter sub-family is
asymptotically flat. In \static~we focused on the one-parameter
sub-family of these in which the aether is aligned with the static
Killing vector. These ``static aether" solutions were found
analytically, up to the inversion of a transcendental function.
The solutions are characterized by their total mass, and it was
shown in \static~that the solution outside a spherical star is in
this family.

In this paper we focus on the time-independent, spherically
symmetric black hole solutions of ae-theory. This is the first
step in comparing ae-theory with astrophysical observations of
black holes. It is also of mathematical interest as an example of
unusual black hole behavior with non-linear self-gravitating
fields.

The Lagrangian for ae-theory depends on four dimensionless
coupling coefficients $c_{1,2,3,4}$. The static aether solutions
referred to above depend only on the single combination $c_1+c_4$,
but this is not the case for the black hole solutions. Moreover,
unlike those solutions, the black hole solutions cannot (as far as
we know) be obtained analytically, so all of the results in this
paper are numerically obtained. We do not make an exhaustive study
here for all values of the $c_i$, but rather just attempt to
determine the generic behavior of the black hole solutions.

We begin this paper in section \ref{generalities} with a
discussion of the definition of a black hole in ae-theory, and the
conditions for the existence of regular black hole solutions. The
qualitative reasons for the existence of a one parameter family of
such solutions are explained. In preparation for the subsequent
detailed analysis the action, field equations, and field
redefinition properties of the theory are reviewed in section
\ref{aebasics}. This is followed in section \ref{Behavior} by a
demonstration using the power series solution of the field
equations about a metric horizon that such horizons are
generically regular, i.e. there is a three parameter family of
solutions in the neighborhood of such a horizon, just as in the
neighborhood of any generic point. We then show that spin-0
horizons are generically singular, but a two-parameter family of
solutions is regular and the additional condition of asymptotic
flatness further reduces this to a one-parameter family. In
section \ref{blackholes} we study the properties of typical
examples of this one parameter family of black holes, imposing
regularity by a power series expansion about the spin-0 horizon
and numerically integrating out to infinity, determining the
asymptotically flat solutions by tuning the data at the horizon.
These black holes are rather similar to Schwarzschild outside the
horizon, and like Schwarzschild they have a spacelike curvature
singularity inside at (or very near) zero radius. Unlike the
static aether solutions of \static, $u^a$ is not aligned with the
static Killing field in these black hole solutions. The aether
flows into the black hole, but differs significantly from the
4-velocity of freely-falling geodesics at rest at infinity. We
also note that some functions constructed from the metric and
aether exhibit oscillatory behavior in the interior as they
approach the singularity. Section \ref{discussion} concludes the
paper with a brief discussion of questions for further work.

\section{General properties of black holes
in ae-theory}

\label{generalities}

The relevant notion of a black hole in ae-theory is not immediately
clear. To trap matter influences a black hole must have a horizon
with respect to the causal structure of $g_{ab}$, the metric to
which  matter couples universally (or almost universally, according
to observations). We call this a ``metric horizon".  This is not the
only relevant notion of causality however. For general values of the
coupling coefficients $c_i$, ae-theory has multiple characteristic
hypersurfaces. In particular, perturbing around flat spacetime it
was found in \cite{Jacobson:2004ts} that there are spin-2, spin-1,
and spin-0 wave modes, with squared speeds relative to the aether
given by
\beq\label{speeds} \eqalign{\mbox{spin-2}\qquad 1/(1-c_{13}) \cr
\mbox{spin-1}\qquad (c_1-\half c_1^2 +\half c_3^2)/c_{14}(1-c_{13})
\cr \mbox{spin-0}\qquad
c_{123}(2-c_{14})/c_{14}(1-c_{13})(2+c_{13}+3c_2)} \eeq
%
where $c_{13}=c_1+c_3$, etc. The speeds $s_i$ are generally
different from each other and from the metric speed of light 1.
Only in the special case where $c_4=0$, $c_3=-c_1$, and
$c_2=c_1/(1-2c_1)$ do all the modes propagate at the same speed.
In general, the characteristic surfaces for a mode of speed $s_i$ are null
with respect to the effective metric
$\eta_{ab}+(s_i^2-1)\underline{u}_a \underline{u}_b$, where
$\eta_{ab}$ is the flat metric and $\underline{u}_a$ the constant
background aether.

In the nonlinear case characteristics can be defined as
hypersurfaces across which the field equations admit a
discontinuity
in first derivatives~\cite{CourantHilbert}. For this paper we presume
that these characteristics define the relevant notion of causal
domain of dependence for the ae-theory field equations. This seems
quite plausible, although no rigorous study has been attempted.
The characteristic hypersurfaces are determined by the highest
derivative terms in the field equations, so can also be identified
by examination of high frequency solutions to the linearized
equations about a given background~\cite{CourantHilbert}. For such
solutions the gradients in the background are irrelevant, so we
can infer that the nonlinear characteristics are null surfaces of
the effective metrics,
\beq g^{(i)}_{ab} = g_{ab}+(s_i^2-1) u_a u_b. \label{geff} \eeq
We refer to the horizons associated with these metrics as the
spin-0, spin-1, and spin-2 horizons. If a black hole is to be a
region that traps all possible causal influences, it must be bounded
by a horizon corresponding to the fastest speed. The coupling
coefficients $c_i$ determine which speed is the fastest.

A horizon is potentially a location where a solution to the field
equation can become singular. This is because at a characteristic
surface the coefficient of a second derivative term usually present
in the equation vanishes. As such a surface is approached, the
smallness of that coefficient may generically produce a solution in
which some second derivative grows without bound, leading to
singular behavior. This does not occur at spin-1 and spin-2 horizons
in spherically symmetric solutions to ae-theory, presumably since
there are no spherically symmetric spin-1 or spin-2 modes. However
there is a spherical spin-0 mode, and we find that spin-0 horizons
are generically singular.

The requirement that the spin-0 horizon be regular reduces the
three parameter family of local static, spherically symmetric
solutions to a two parameter family, which reduces to a one
parameter family when asymptotic flatness is imposed. Hence there
is just a one parameter family of regular static, spherically
symmetric black hole solutions in ae-theory, just as in GR.
Unlike outside a star, the
aether in these solutions is {\it not} aligned with the Killing
vector but rather flows into the black hole.
We conjecture that the collapse of a spherical star to
form a black hole is nonsingular, and therefore must be
accompanied by a burst of spin-0 aether radiation as the aether
(and metric) exterior adjusts.

The fact that the aether is not aligned with the Killing vector in
a static black hole solution is no accident. It cannot be so
aligned, since it is everywhere a timelike unit vector and the
Killing vector is null on the horizon. Thus at a regular horizon
the aether must be ``infalling", although it is nevertheless
invariant under the Killing flow. The static aether solution found
in \static~has aether aligned with the Killing vector, and can be
thought of as an extremal black hole with a singular horizon on
which the aether becomes infinitely stretched.

While the aether can be regular at a generic point on the horizon,
it cannot smoothly extend to the bifurcation sphere $\cal B$, i.e
the fixed point set of the Killing flow at the intersection of the
past and future horizons~\cite{Eling:2004dk}.  The Killing flow
acts as a Lorentz boost in the tangent space of any point on $\cal
B$, so it is impossible for the aether to be invariant under the
flow there. This implies that the aether must blow up, becoming an
infinite null vector as $\cal B$ is approached. This in turn
raises the concern that there may be no regular metric horizon,
since regularity on a future horizon is typically linked to
regularity at $\cal B$. Indeed Racz and Wald~\cite{Racz:1995nh}
have established, independent of any field equations, conditions
under which a stationary spacetime with regular Killing horizon
can be extended to a spacetime with a regular bifurcation surface,
and conditions under which matter fields invariant under the
Killing symmetry can also be extended. In spherical symmetry these
conditions are satisfied for the metric, but the aether vector
field breaks the required time reflection symmetry so it need not
be regular at the bifurcation surface (although all scalar
invariants must be, as must the aether stress tensor if the field
equations hold).

Another potential obstruction to the existence of regular ae-theory
black hole horizons arises from the form of the aether stress
tensor. At a regular stationary metric horizon the Raychaudhuri
equation for the horizon congruence with with null generator $k^a$
implies that $R_{ab}k^a k^b$ must vanish, hence the Einstein
equation implies that the matter stress tensor component
$T_{ab}k^a k^b$ must also vanish. With common matter fields, e.g.,
scalar fields, Maxwell or Yang-Mills fields, and nonlinear sigma
model fields, it is easy to show from examination of the form of
the stress tensors that this condition is automatically satisfied
locally for any field invariant under the Killing flow,
independent of field equations. This property does not seem to
hold kinematically for the aether stress tensor, but since we find
a full three-parameter family of regular metric horizons,  it is
evidently imposed  by the field equations.

The fact that $T_{ab}k^ak^b$ does not vanish kinematically might
appear to contradict the following general argument. For a Killing
horizon with non-zero surface gravity it is not necessary to examine
the form of the stress tensor to arrive at the inference that the
horizon component of a matter stress tensor vanishes. If $\chi^a$ is
the horizon-generating Killing vector, then the vanishing of the
scalar $T_{ab}\chi^a\chi^b$ on the horizon is guaranteed by the
facts that (i) it is invariant along the flow, (ii) $\chi^a$
vanishes at the bifurcation surface, and (iii) $T_{ab}$ is regular
at the bifurcation surface (as guaranteed by the Racz-Wald extension
theorem). But this argument too seems to fail for the aether stress
tensor, because (as noted above) there is no purely kinematic way to
argue that it (and therefore its stress-tensor) is regular at the
bifurcation surface. So, again, the field equations seem to play an
essential role in ensuring the existence of regular metric horizons.

\section{Einstein-Aether theory}
\label{aebasics}

The action for Einstein-Aether theory is the most general
diffeomorphism invariant functional of the spacetime metric
$g_{ab}$ and aether field $u^a$ involving no more than two
derivatives,
\beq S = \frac{1}{16\pi G}\int \sqrt{-g}~ L ~d^{4}x
\label{action} \eeq
where
\beq L = -R-K^{ab}{}_{mn} \nabla_a u^m \nabla_b u^n -
\lambda(g_{ab}u^a u^b - 1). \eeq
Here $R$ is the Ricci scalar,  $K^{ab}{}_{mn}$ is defined as
\beq K^{ab}{}_{mn} = c_1
g^{ab}g_{mn}+c_2\delta^{a}_{m}\delta^{b}_{n}
+c_3\delta^{a}_{n}\delta^{b}_{m} +c_4u^au^bg_{mn} \eeq
where the $c_i$ are dimensionless constants, and $\lambda$ is a
Lagrange multiplier enforcing the unit timelike constraint. This
constraint restricts variations of the aether to be spacelike, hence
ghosts need not arise. A term of the form $R_{ab} u^a u^b$ is not
explicitly included as it is proportional to the difference of the
$c_2$ and $c_3$ terms in (\ref{action}) via integration by parts.
The metric signature is $({+}{-}{-}{-})$ and the units are chosen so
that the speed of light defined by the metric $g_{ab}$ is unity.
When $u^a$ is a unit vector, the square of the twist
$\omega_a = \epsilon_{abcd} u^b \nabla^c u^d$ is a combination
of the $c_1$, $c_3$, and $c_4$ terms in the action ({\ref{action}),
\beq \omega_a \omega^a = -(\nabla_a u_b)(\nabla^a u^b) + (\nabla_a
u_b)(\nabla^b u^a)+ (u^b \nabla_b u_a)(u^c \nabla_c u^a).\eeq
In spherical symmetry the aether is hypersurface orthogonal, hence it
has vanishing  twist, so
the $c_4$ term can be absorbed by making the replacements
\beq
c_1
\rightarrow c_1 + c_4,\qquad c_3 \rightarrow c_3 - c_4,\qquad c_4\rightarrow0.
\label{c4absorb}\eeq

The field equations from varying (\ref{action}) plus a matter action
(coupled only to the metric) with respect to $g^{ab}$, $u^a$ and
$\lambda$ are given by
\begin{eqnarray}
G_{ab} = T^{(u)}_{ab}+8\pi G T^{M}{}_{ab}\label{AEE}\\
\nab_a J^{a}{}_m-c_4 \dot{u}_a \nab_m u^a = \l u_m,
\label{ueqn}\\
g_{ab} u^a u^b = 1, \label{constraint}
\end{eqnarray}
where
\beq J^a{}_{m} = K^{ab}{}_{mn} \nabla_b u^n. \label{Jdef}\eeq
The aether stress tensor is given by
\bea T^{(u)}{}_{ab}&=&\nab_m(J_{(a}{}^m u_{b)}-J^m{}_{(a} u_{b)}
- J_{(ab)}u^m) \nonumber\\ &&+ c_1\, \left[(\nab_m u_a)(\nab^m u_b)-(\nab_a u_m)(\nab_b
u^m) \right]\nonumber\\ &&+c_4\, \dot{u}_a\dot{u}_b\nonumber\\
&&+\left[u_n(\nab_m J^{mn})-c_4\dot{u}^2\right]u_a u_b \nonumber\\
&&-\frac{1}{2} L_u g_{ab}, \label{aetherT}\eea
where $L_{u} = -K^{ab}{}_{mn} \nabla_a u^m \nabla_b u^n$. The
Lagrange multiplier $\lambda$ has been eliminated from
(\ref{aetherT}) by solving for it via the contraction of the aether
field equation (\ref{ueqn}) with $u^a$.

\subsection{Metric redefinitions}

It is sometimes convenient to re-express the theory in terms of a
new metric and aether field, related to the original fields by a
field redefinition of the form
\begin{eqnarray}
g'_{ab} = g_{ab}+(\s-1) u_a u_b \label{redef1}\\
u'^a = \frac{1}{\sqrt{\s}} u^a. \label{redef2}
\end{eqnarray}
The constant $\s$ is restricted to be positive so the new metric
remains Lorentzian. In effect, the field redefinition ``stretches"
the metric tensor in the aether direction by a factor $\s$. The action
({\ref{action}) for ($g'_{ab}$, $u'^{a}$) takes the same form as
that for ($g_{ab}$, $u^{a}$) up to the values of the coefficients
$c_i$. The $c_i$ coefficients for the action expressed in terms of the
new fields (\ref{redef1})-(\ref{redef2}) are~\cite{Foster:2005ec}
\bea c'_1 = \frac{\s}{2}\Bigl((1+\s^{-2})c_1+(1-\s^{-2})c_3-
(1-\s^{-1})^2\Bigr)\label{fd1}\\
c'_2 =\s(c_2+1-\s^{-1})\label{fd2}\\
c'_3 = \frac{\s}{2}\Bigl((1-\s^{-2})c_1+(1+\s^{-2})c_3-
(1-\s^{-2})\Bigr)\label{fd3}\\
c'_4 = c_4 - \frac{\s}{2}\Bigl((1-\s^{-1})^2c_1+(1-\s^{-2})c_3-
(1-\s^{-1})^2\Bigr) \label{fd4}. \eea
%
Certain combinations of the coefficients change by a simple
scaling under the field redefinitions:
\bea
c'_{14} = c_{14}\\
c'_{123} = \s c_{123}\\
c'_{13}-1 = \s (c_{13}-1)\\
c'_1-c'_3-1 = \s^{-1}(c_1-c_3-1). \eea
The field redefinition relates solutions to the field
equations coming from the two actions.

Using a field redefinition the general form of the action can be
simplified~\cite{Foster:2005ec} by eliminating one of the $c_i$ or
a combination of the $c_i$. If we choose $\s = (s_2)^2=
1/(1-c_{13})$, then $c'_{13}$ vanishes, i.e. $c'_3=-c'_1$. In this
case the two corresponding terms in the Lagrangian combine to make
a Maxwell-like lagrangian, for which the Levi-Civita connection
drops out. (This choice of $\s$ is positive and therefore
preserves Lorentzian signature provided
the original coefficients satisfy $c_{13}<1$.) In the context of
spherical symmetry, we may also exploit the vanishing to the twist
of the aether to absorb the $c_4$ term in ({\ref{action}) by the
replacements (\ref{c4absorb}) as explained above. After these two
changes, the Lagrangian takes a much simpler reduced form
characterized by only two $c_i$ coefficients,
\beq L_{\rm ae} = -R - \frac{c'_1}{2} F^{ab} F_{ab} - c'_2
(\nabla_a u^a)^2 - \l(g_{ab} u^a u^b -1), \label{ssredaction}\eeq
where $F_{ab} = 2 \nabla_{[a} u_{b]}$. In the next section we will
use this reduced form of the action to investigate the general
behavior of stationary spherically symmetric solutions, addressing
the existence of solutions around a metric horizon, asymptotic
flatness, and the regularity of the spin-0 horizon.

Another useful
choice of field redefinition
is to arrange for the new metric in (\ref{redef1}) to
coincide with the effective metric for one of the wave modes
(\ref{geff}) by choosing $\s=s_i^2$. In this way we transform to a
``frame" where one of the spin-2, spin-1, or spin-0 horizons
coincides with the metric horizon. Under the field redefinition
all the squared  speeds become $s'_i{}^2=s_i^2/\s$, since the
metric tensor in the aether direction
is stretched by the factor $\s$. In
section \ref{blackholes}  we use this method to
make the spin-0 and metric horizons coincide,
which simplifies the expansion of the field equations
around the spin-0 horizon.

\section{Generic behavior of horizons and spatial infinity}
\label{Behavior}

In this section we demonstrate that, for generic values of the
$c_i$ coefficients (at least for the reduced theory
(\ref{ssredaction})), (i) metric horizons are generically regular
in the three-parameter family of local solutions, (ii) asymptotic
flatness imposes one condition on this family, and (iii)
regularity of the spin-0 horizon imposes another condition,
leaving a one-parameter family of black hole solutions with
regular spin-0 horizons.

\subsection{Metric horizon expansion}
\label{metrichor}

One way to determine the number of independent solutions with a
regular metric horizon is to expand the field equations in a power
series about a candidate horizon and solve algebraically order by
order. We did this using the Maple computer application to carry
out the algebra. The computation was prohibitively complicated
using our methods with general values of the $c_i$, so we
restricted attention to the reduced theory (\ref{ssredaction}).
The determination of the solution space can also be done more
``experimentally," by numerical integration of the field equations
with varying initial data. With the latter method no special
restriction on the $c_i$ is required.

For the study of spherical black holes we adopt
Eddington-Finkelstein (EF) type coordinates $(v,r,\theta,\phi)$ with
line element
\beq ds^2 = N(r) dv^2 - 2B(r) dv dr - r^2 d\Omega^2 \label{EF}\eeq
with advanced (null) time coordinate $v$ and radial ``area
coordinate" $r$. The time-translation Killing vector is
$\partial_v$, and a metric horizon corresponds to $N(r)=0$. These
coordinates are regular at metric and other horizons so are useful
for studying black holes and their interiors. Using these
coordinates a stationary spherical aether field takes the form
\beq u^a =  a(r)\partial_v+ b(r)\partial_r\label{uform}\eeq
and the unit constraint (\ref{constraint}) becomes
\beq Na^2-2 Bab = 1. \label{constr1} \eeq
The field equations (\ref{AEE}) and (\ref{ueqn}) become a set of
coupled, second order ordinary differential equations (ODE's) involving
the functions $N, B, a, b$. We use the
constraint (\ref{constr1}) to solve for $b$ in terms of the other
three functions. Even for the reduced case $c_{13}=c_4=0$ the
equations are sufficiently complicated that it does not seem useful
to display them here.

Regularity at the metric horizon can be imposed by making a power
series expansion about the radius $r_h$ where $N(r_h)= 0$,
\bea
N(r) &= N'(r_h)(r-r_h) + \half N''(r_h)(r-r_h)^2 + \cdots \label{Nhor}\\
B(r) &= B(r_h) + B'(r_h)(r-r_h) + \half B''(r_h)(r-r_h)^2 + \cdots \label{Bhor}\\
a(r) &= a(r_h) + a'(r_h)(r-r_h) + \half a''(r_h)(r-r_h)^2 + \cdots
\label{ahor} \eea
Inserting these expansions in the field equations, one can solve
order by order for the power series coefficients. This allows the
set of free parameters in the initial data at the horizon to be
identified. At zeroth order in $(r-r_h)$ the field equations imply
that $a'(r_h)$ is a function of $N'(r_h)$, $B(r_h)$, $a(r_h)$, and
$r_h$. The specific result is sufficiently complicated that it too
does not seem useful to display here. Solving to higher orders we
find that all remaining coefficients in the series expansion are
determined by these four initial data parameters.  Using the
scaling freedom in the $v$ coordinate ($v \rightarrow \l v$ where
$\l$ is a constant) one of the initial values at the horizon can
be fixed arbitrarily. Thus, there is a three-parameter family of
local solutions with a regular metric horizon. As discussed in
\static, we also found a three-parameter family of local solutions
expanding about an arbitrary radius (i.e. not imposing a horizon),
hence we conclude that regularity of the metric horizon
generically imposes no restriction on the solutions.

\subsection{Asymptotic expansion}
\label{asymexp}

In addition to regularity at all the horizons, we require the
black hole solutions to be asymptotically flat. To determine the
form of such solutions we can change to the inverse radius
variable $x=1/r$ and expand around $x=0$, as was done in
\cite{Eling:2003rd} (where isotropic coordinates were employed).
In the reduced theory (\ref{ssredaction}) this yields the
solutions
\begin{eqnarray}
N(x) &= 1+N_1x+\frac{1}{48} c'_1 N_1^3 x^3+ \cdots \label{asyN}\\
B(x) &= 1+\frac{1}{16} c'_1 N_1^2 x^2-\frac{1}{12} c'_1 N_1^3
x^3+\cdots \label{asyB} \\
a(x) &= 1-\half N_1 x+a_2 x^2+(-\frac{1}{96} c'_1
N_1^3+\frac{1}{16} N_1^3-N_1 a_2) x^3+\cdots \label{asya}
\end{eqnarray}
where $N_1=N'(x=0)$ and $a_2 = a''(x=0)$,
and the freedom to rescale $v$ has been exploited
to set $N(x=0)=1$.
No more free parameters appear at higher orders, so
the asymptotically flat solutions are determined by
the two free parameters $N_1$ and $a_2$.

An asymptotically flat solution can be determined by using a
simple shooting method, numerically integrating outward from an
interior radius where there are three free initial data
parameters. As in \static, we find that to match the asymptotic
form (\ref{asyN})-(\ref{asya}) requires tuning just one of the
three initial parameters, as expected since there are two free
parameters in the asymptotic form.  We conclude that in particular
there is a two-parameter family of asymptotically flat ``black
hole" solutions with metric horizon fixed to lie at a given radius
$r_h$. In practice, to integrate outward from a metric horizon we
found it necessary to first use the perturbative solution about
the horizon to generate from the horizon data an initial data set
some small radial distance away. This is because the ODE's have a
singular point at the horizon.

A more direct way to generate such asymptotically flat black hole
solutions is to start the numerical integration near infinity and
integrate inward using the inverse radius coordinate $x$. Since
$x=0$ is a singular point of the ODE, it is necessary to start the
integration at some small non-zero $x$ value. The expansions
(\ref{asyN})-(\ref{asya}) can be used to generate initial data as a
function of $N_1$ and $a_2$. We find again this way that regularity
at the metric horizon does not impose any conditions on $N_1$ and $a_2$.
The functions $N$, $B$, and $a$ evolve smoothly through a point where
$N$ goes to zero.

\subsection{Asymptotically flat solutions and spin-0 horizon regularity}

So far we have shown that there is a two-parameter family of
asymptotically flat black hole solutions with a regular metric
horizon. Normally we expect just one black hole parameter, the total
mass, unless there are conserved charges that can be additional
parameters. In ae-theory there seems to be no such conserved charge,
so the situation is puzzling. Another puzzling aspect of these black
hole solutions not yet discussed here is that some have internal
singularities at nonzero radius, rather than just at $r=0$ like most
known black holes. Moreover, in some solutions we found that such
singularities can occur externally, i.e. not inside a metric horizon. All these
puzzles are resolved by the recognition that the singularities in
question occur precisely at the location of the spin-0 horizon.
Imposing regularity at the spin-0 horizon eliminates one free
parameter, leaving us with a conventional one-parameter family of
asymptotically flat black holes.

In the rest of this subsection the full range of behavior of
asymptotically flat solutions is discussed. In particular it is
demonstrated that when a spin-0 horizon occurs it is singular for
generic values of the two initial data parameters $N_1$ and $a_2$ at
infinity. Evidence is given that by tuning one of these parameters
to a special value a regular spin-0 horizon can be obtained. In the
next section we show by a power series expansion around the spin-0
horizon that such regular solutions do indeed exist.

For the reduced theory (\ref{ssredaction}), the spin-2 and spin-1
speeds are both unity, but the spin-0 mode has squared speed
$s_0^2=(c'_2/c'_1)(2-c'_1)/(2+3c'_2)$ (relative to the aether),
which is generically different from unity. At the spin-0 horizon the
mode propagates at fixed radius, hence the surface of constant $r$
at that location is null with respect to the effective metric
$g^{(0)}_{ab}$ defined in (\ref{geff}). This implies the condition
$g^{(0)}_{vv}=0$ in EF coordinates, which we find occurs where
\beq N a^2 = \frac{1-s_0}{1+s_0}\quad\mbox{or}\quad
\frac{1+s_0}{1-s_0}. \label{horcond} \eeq
The first root is less than 1 so according to (\ref{constr1}) occurs when
$b<0$, i.e. when the aether tips inward. For the second root the aether
tips outward.
The combination $f=Na^2$ is independent of the arbitrary scale for
the $v$ coordinate and equal to one at infinity. Inserting the
expansions (\ref{asyN}) and (\ref{asya}) we find the asymptotic form
\beq f(x)=N(x)a(x)^2=1+2(a_2-\frac{3}{8}N_1^2)x^2-N_1
(a_2-\frac{3}{8}N_1^2)x^3+O(x^4)\eeq
Curiously, this expansion is independent of both $c'_1$ and $c'_2$
through order $x^3$ (but not beyond) and depends linearly on $a_2-
\frac{3}{8}N_1^2$ through order $x^5$. This pattern suggests an
analytic solution may be possible, but we will not pursue this
here. The static aether solution studied in \static~corresponds to
the case where $f(x)=1$ for all $x$, which occurs when
$a_2/N_1^2=3/8$.

As in the previous subsection, we study solutions obtained by
integration inwards starting from an asymptotically flat spatial
infinity. Since the theory has no length scale, the solution with
data $(N_1,a_2)$ is trivially related to that with data $(\lambda
N_1,\lambda^2 a_2)$ (as are Schwarzschild solutions with different
mass trivially related). If we think of the line element $ds^2$ as
giving a numerical value specified with respect to a given length
unit, then to go from one solution to another we need only change
the unit of length. Thus without loss of generality we can fix units
with $N_1=\pm1$. The solutions then depend on the choice of theory
through $c'_1$, $c'_2$, on the parameter $a_2$, and on the sign of
$N_1$. A systematic study of these solutions is beyond the scope of
this paper; here we just indicate the various behaviors we have
encountered, and then focus on the regular positive mass black
holes.

Let us first consider positive mass solutions, i.e. $N_1=-1$. As the
radial coordinate decreases, $N$ decreases from 1, while the
combination $Na^2$ increases or decreases according as $a_2$ is
greater or less than 3/8.
There are solutions where $f(r)$ does not reach 0 or $(1\mp
s_0)/(1\pm s_0)$ and therefore neither a metric nor spin-0 horizon
is attained. In some of these $N(r)$ re-curves out to positive
infinity, $a(r)$ approaches zero and the solution reaches a
curvature singularity near $r=0$. There are also solutions similar
to the static aether of \static, where $B(r)$ goes to infinity as
$N(r)$ approaches a finite value, indicating a minimal area
two-sphere. In some cases the larger root for a spin zero horizon
in (\ref{horcond}) is reached by $f(r)$. In other solutions $f(r)$
may reach a metric horizon, but does not attain the value
corresponding to a spin-0 horizon. This can only happen when $s_0
> 1$. $N(r)$ again re-curves out to infinity and $a(r)$ approaches
zero near $r=0$ and there are outer and inner metric horizons. In
still other solutions, the functions and their derivatives are
regular up to the point where $f(r)$ reaches the spin-0 horizon, but
generically the spacetime is singular at that point. If $s_0 > 1$,
$f(r_h)$ is negative and the singularity is located inside a metric
horizon. If $s_0 < 1$ the singularity occurs without any metric
horizon.

For a specific example of this last type we choose parameters
$c'_1=0.051$ and $c'_2=0.116$, for which the spin-0 speed is 1.37.
(These arise from starting with coefficients that satisfy all the
observational constraints described in \cite{Foster:2005dk} and
performing the field redefinition to the reduced action
(\ref{ssredaction}).) Figure \ref{generic}
\begin{figure}
\begin{center}
\includegraphics[angle=270,width=12cm]{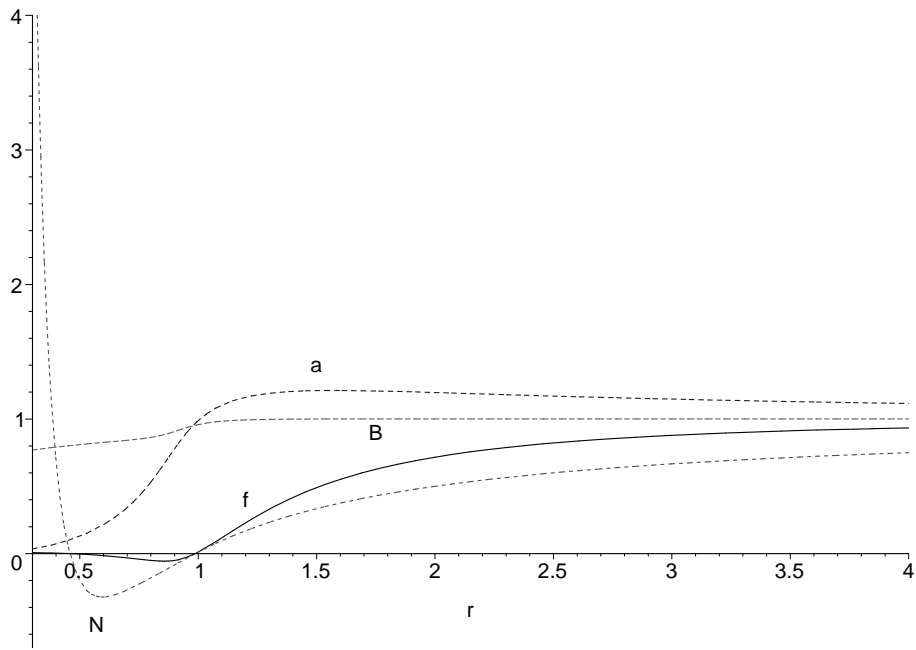}\\
\caption{\label{generic}Solution for reduced theory
(\ref{ssredaction}) with $c'_1=0.051$ and $c'_2=0.116$, determined
by data $N_1=-1$ and $a_2=-0.1$ at spatial infinity. There is both
an outer and inner metric horizon where $N$ vanishes, but $f$ does
not decrease enough to reach a spin-0 horizon. The functions $N$ and
$f$ go to zero slightly inside $r=1$ which would be the horizon
radius of the corresponding Schwarzschild solution. As $a_2$
increases the minimum of $f(r)$ decreases until the solution
acquires a spin-0 horizon, where is it generically singular.}
\end{center}
\end{figure}
shows the behavior of $N$, $B$, and $a$ for $N_1=-1$ and
$a_2=-0.1$. In this case there are ``outer" and ``inner" metric
horizons where $N=0$, but $f(r)$ does not reach as low as
$-0.158$, which is required by (\ref{horcond})  in this case for a spin-0 horizon. For
values of $a_2< -0.1$ the minimum value of $N$ shifts upward and
eventually $N$ never reaches zero, i.e. the metric horizon
disappears.  On the other hand, for $a_2>
-0.1$ the minimum value of $f$ decreases until the spin-0 horizon
is reached. At this point $a'$ goes to negative infinity, $N'$
blows up to positive infinity, and there is a curvature
singularity. In contrast, note that in Figure \ref{generic}
$a'(r)$ has a maximum value while $N'(r)$ goes to {\it negative}
infinity. This suggests that at some special value of $a_2$ there
is a transition were the concavity of $a'(r)$ and $N'(r)$ changes
and the derivatives are finite at a spin-0 horizon. Regularity at
the spin-0 horizon seems thus to impose one condition on the
asymptotic values $N_1$ and $a_2$.

In the negative mass case one might expect only solutions
analogous to negative mass Schwarzschild, with $N$ increasing from
1 at spatial infinity and no spin-0 or metric horizon. While the
solution does take this form for all $a_2$ in the theory with
$c'_1=0.051$, $c'_2=0.116$, for other values of $c'_{1,2}$
 there are ranges of
$a_2$ where $N$ increases from 1 at infinity, but then decreases
to a metric horizon at finite $r$, and all the functions and their
derivatives are regular until $f(r)$ reaches the value less than
zero required for a spin-0 horizon. This peculiar behavior of a
negative mass solution with metric and spin-0 horizons remains to
be studied more closely. In particular, it is not clear whether a
negative mass solution with a regular spin-0 horizon could exist.

\section{Black holes with regular spin-0 horizons}
\label{blackholes}

In this section we discuss the behavior of black hole solutions
possessing regular spin-0 horizons. Rather than imposing
regularity at the spin-0 horizon by the shooting method integrating
in from infinity, we instead expand the field equations in a power
series about a non-singular spin-0 horizon.

\subsection{Horizon expansion}
\label{bh:horexp}

Due to the complexity of the field equations and their singular
nature at the horizon, we were unable to implement the power
series solution about a spin-0 horizon in the generic reduced
theory (\ref{ssredaction}) (even with computer aided algebra). It
might be possible to obtain the perturbative solution by a more
well-adapted method, but instead we simplified the computation by
making a field redefinition to a new metric for which the spin-0
and metric horizons coincide. Starting from an arbitrary set of
coefficients $c_i$, this is implemented by the choice $\s=s_0^2$
in (\ref{redef2}), after which we have $s_0=1$ without loss of
generality in the theory. As before we can also then exploit
spherical symmetry to absorb $c_4$ by making the replacements
(\ref{c4absorb}), which do not disrupt the coincidence of the
spin-0 and metric horizons since this is just a re-expression of
the same Lagrangian without changing the field variables
\footnote{The spin-0 speed is invariant under (\ref{c4absorb}), as
guaranteed by this argument. The spin-1 speed is {\it not}
invariant, but this does not contradict the argument since there
is no spherically symmetric spin-1 mode. Note however that in
diagnosing whether spin-1 perturbations are trapped in a given
black hole it is important to use the value of the spin-1 speed
written in (\ref{speeds}) {\it before} the $c_4$ coefficient has
been absorbed.}. This reduces the distinct parameter space to just
($c_1$,$c_3$) (omitting the prime in the notation for $c_{1,3}$).
After this field redefinition the coefficient $c_2$ is given by
\beq c_2 =
\frac{-2c_3-c_1^3-2c_3c_1^2-c_1c_3^2}{2-4c_1+3c_1^2+3c_3c_1}.
\label{c2}\eeq

A further simplification of the equations is achieved by trading
the metric function $N$ for the combination of metric and aether
functions $f= Na^2$. We have no insight into why this simplifies
the expansion of the field equations about the common metric and
spin-0 horizon at $N=0=f$, although as stated above the
combination $Na^2$ is invariant under a rescaling of the $v$
coordinate. The field equations in this set of field variables
involve $a$, $a'$, $a''$, $f$, $f'$, $f''$, $B$, and $B'$. At the
horizon $f(r)$ vanishes linearly, $f(r) = f'(r_0)(r-r_0)+ \cdots $
By a constant rescaling of $v$ we can furthermore set $B(r_h)$
equal to 1. Using this along with (\ref{Bhor}) and (\ref{ahor})
the field equations can be expanded and solved order by order for
the coefficients of the power series.

Solving the field equations for this theory as algebraic equations
for the expansion coefficients we find that at zeroth order in
$(r-r_h)$ the quantities $a(r_h)$, $a'(r_h)$, $a''(r_h)$, and
$f''(r_h)$ are determined by free parameters $r_h$, $f'(r_h)$,
$B'(r_h)$.   We succeeded in solving the equations to the next
order in $(r-r_h)$ only in the special cases $c_3=0$, $c_3=c_1$,
and $c_3=-c_1$. In these cases we find that $B'(r_h)$ is
determined by $r_h$ and $f'(r_h)$. Hence, consistent with the
expectation of the previous section, there is a two-parameter
family of local solutions around the regular spin-0 horizon. These
solutions are generically not asymptotically flat.

\subsection{Asymptotically flat black holes}
\label{bh:asymflat}

To produce asymptotically flat solutions we numerically integrate
outward, starting with the horizon data and matching onto
(\ref{asyN}), (\ref{asyB}), and (\ref{asya}) by tuning $f'(r_h)$
until $f(r)$ is constant and equal to 1 at very large $r$ values.
The asymptotic flatness boundary condition at infinity thus reduces
the number of free parameters to one, namely the horizon radius
itself. Solutions with different horizon radii are trivially
related. Since $r_h$ is a singular point of the ODE's, it is
necessary to start the integration with initial data at some small
positive value of $r-r_h$. We used the series solution determined by
a given $r_h$ and $f'(r_h)$ to generate this initial data. To
examine the solution inside the horizon, we numerically integrated
inward, starting at a small negative value of $r-r_h$ with data
generated by the same series solution.

Here we will discuss the
properties of the solutions for the $c_3=0$ theory only, whose
behavior is typical of the three special cases $c_3=0,\pm c_1$.
Figure \ref{bhsol1} displays the solution for $c_1=0.3$,
$c_2=-.025$, $c_3=c_4=0$,  together
with $S(r)=1-2/r$, the Schwarzschild version of $N(r)$ with the
same mass.
\begin{figure}
\begin{center}
 \includegraphics[angle=270,width=12cm]{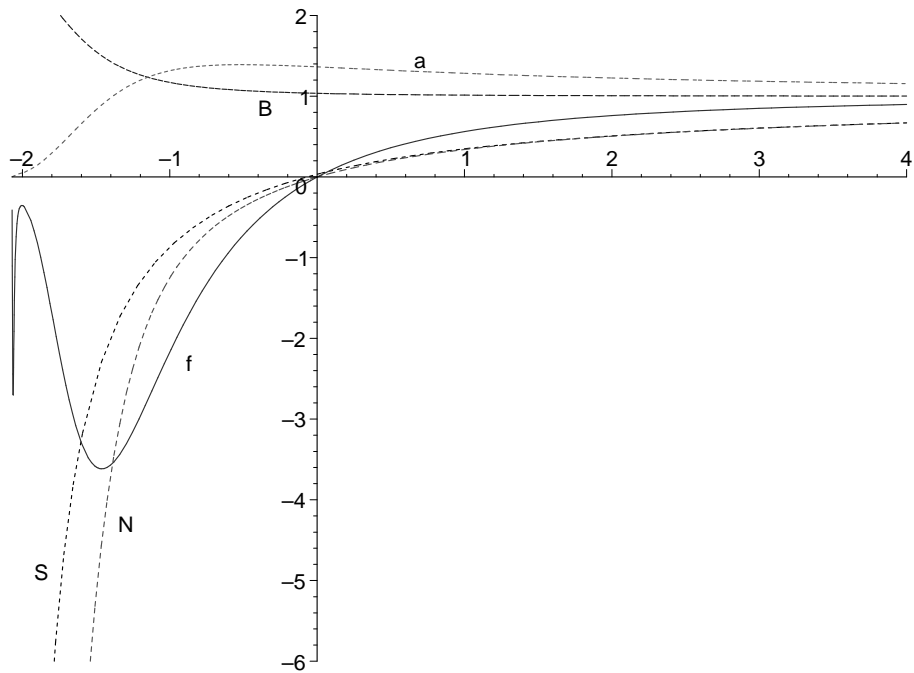}\\
\caption{\label{bhsol1}Plots of $f$, $N$, $B$, $a$, and $S$ (the
Schwarzschild version of $N$) vs. $z=r-r_h$ for $c_1=0.3$, in units
with $r_0=2$. The horizon radius is $r_h\approx 2.07$ for the
ae-theory black hole, so $S=0$ at $z\approx-0.07$. The solutions
agree closely outside the horizon.  Deviations are noticeable near
the horizon and become significant in the interior, where $N$ blows
up more rapidly. Near the singularity $f$ begins to oscillate
rapidly.}
 \end{center}
\end{figure}
For this plot we use the scaling freedom of
$v$ to convert the numerical solution to a ``gauge" where the
metric functions and $v$ component of the aether are all equal to
1 at infinity. The two metric functions $B(r)$ and $N(r)$ in
GR and ae-theory are
in very close agreement outside the horizon, while inside they
differ noticeably.

\subsubsection{Black hole mass}
\label{sec:mass}

The ADM mass $M_{ADM}$ of an asymptotically flat spacetime whose
asymptotic metric takes the Schwarzschild form at $O(1/r)$ is
directly determined by the coefficient $r_0=2GM_{ADM}$ of the
$O(1/r)$ part of $g_{tt}$. In ae-theory the relation between
$M_{ADM}$ and the total energy $E$ of the spacetime is
$GM_{ADM}=G_NE$~\cite{Foster:2005fr, Eling:2005zq},
where $G_N=G/(1-c_{14}/2)$ is the Newton constant
appearing in the force law between two weakly gravitating
masses~\cite{Carroll:2004ai}.  We shall refer to the quantity
$r_0/2$ with dimensions of length as the ``mass" in what follows,
and denote it by $M$.
For a Schwarzschild black hole in GR, $r_0$
is equal to the horizon radius $r_h$. In ae-theory the ratio
$r_0/r_h$ is a constant (since there is only one length scale)
determined by the coupling coefficients $c_i$.

The EF line element (\ref{EF}) transforms to
Schwarzschild form
\beq ds^2 = N\, dt^2 - (B^2/N)\, dr^2 - r^2 d\Omega^2 \label{Schw}\eeq
with time coordinate $t$ defined by $dt=dv - (B/N)dr$. The
asymptotic form of $B$ (\ref{asyB}) shows that $B=1+O(1/r^{2})$, so
up through $O(1/r)$ the line element (\ref{Schw}) has the
standard asymptotic form if  $N$ and $B$ are normalized to 1 at infinity.
In generating an asymptotically flat numerical solution we fixed the
scale freedom of the $v$ coordinate by imposing $B(r_h)=1$ at the
horizon however, so the asymptotic form of $N$ is $N_\infty+N_1/r +
O(1/r^2)$. The mass is given
by $M=r_0/2=N_1/2N_{\infty}$, which can be extracted from the numerical
solution at large $r$.

\subsubsection{Horizons}

The solution displayed in Figure~\ref{bhsol1} has metric and
spin-0 horizons at $z=0$, but how about spin-1 and spin-2
horizons? Is the fastest speed actually trapped? The condition for
a horizon corresponding to a speed $s_0$ is given in
(\ref{horcond}). As the speed approaches infinity the horizon
value of $f$ approaches $-1$ from above. In Figure~\ref{bhsol1}
(and for all values of $c_1$ that we studied up to $0.7$), the
minimum value of $f(r)$ is less than $-1$, which is sufficient to
trap any wave mode. The fact that $f(r)$ curves back to being
greater than $-1$ indicates that an inner horizon might exist for
some wave modes in certain parameter ranges  of $c_i$.

In the theory under discussion we have $c_3=c_4=0$ and
$c_2=-c_1^3/(2-4c_1+3c_1^2)$, so the squared mode speeds in
(\ref{speeds}) are given by $1/(1-c_1)$ for spin-2 and
$(1-c_1/2)/(1-c_1)$ for spin-1. With $0<c_1<1$ both of these are
greater than 1, and the spin-2 speed is the highest. In the
particular case shown in the figure, the spin-1 speed is $1.10$
and the spin-2 speed is $1.20$, which correspond to horizons at
$f=-0.049$ and $f=-0.089$ respectively, which do not seem to be
reached a second time.

\subsubsection{Oscillations}
\label{oscillations}

A notable aspect of the black hole interior displayed in
Figure~\ref{bhsol1} is the oscillation in $f$. The function $h=Ba$
also oscillates in a similar manner, but is 180 degrees out of
phase. In addition, there are related oscillations in the
curvature scalar and aether congruence behavior discussed below.
These oscillations are reminiscent of the interior behavior found
in Einstein-Yang-Mills black holes \cite{Donets:1996ja}, where the
metric functions and derivative of the Yang-Mills potential
oscillate an infinite number of times before the singularity.

While $N$ decreases monotonically and $B$ increases monotonically,
$a$ goes to zero, so the oscillations of $Na^2$ and $Ba$ arise
because of variations in the magnitude of their derivatives. Since
the oscillations inside $z=-2$ are not clearly visible in
Figure~\ref{bhsol1} a zoomed in graph of $f(z)$ is provided in
Figure~\ref{fzoom}.
\begin{figure}
\begin{center}
 \includegraphics[angle=270,width=12cm]{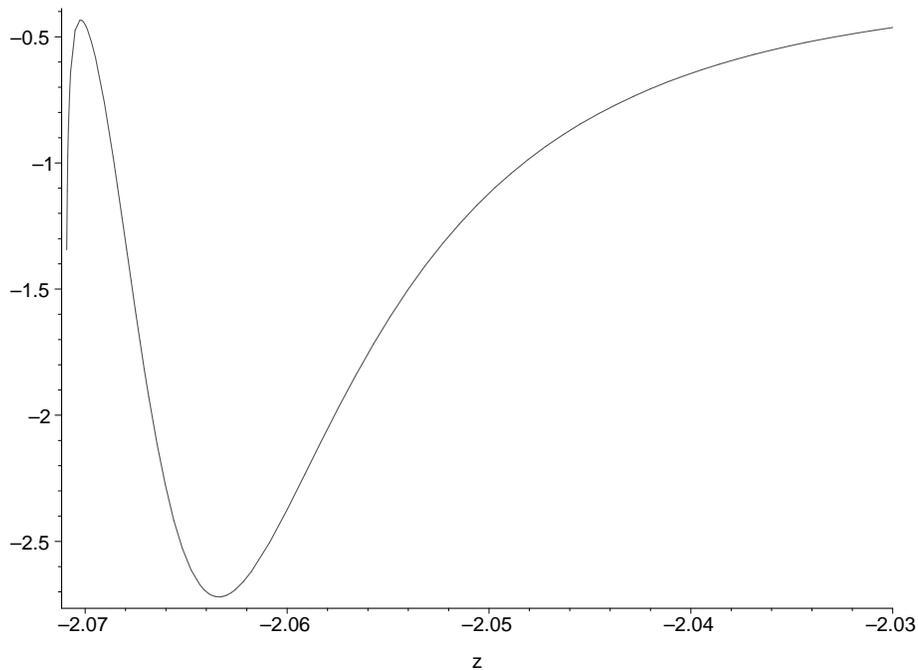}\\
\caption{\label{fzoom}Oscillations of $f=Na^2$ near the singularity
inside the black hole solution shown in figure \ref{bhsol1}, plotted
vs. $z=r-r_h$ in units with $M=1$.}
\end{center}
\end{figure}
{}From this graph it is clear that $f$ smoothly turns over at least
once more before the singularity. Although the number of
oscillations before the singularity appears finite, it is possible
that the numerical integration employed is not capable of resolving
additional or even infinitely more oscillations. More information
may be obtained in the future by improved numeric methods or
analytic methods around $r=0$.

\subsubsection{Curvature singularity}
\label{curvature}

There appears to be a spacelike curvature singularity at or near
$r=0$, as in the Schwarzschild solution of GR. In
Figure~\ref{bhsol1}, the approach of $N$ to negative infinity near
$r=0$  suggests a singularity. In Figure \ref{fig:curvature} the
logarithm of the Kretschmann scalar $K=R_{abcd} R^{abcd}$ is plotted
vs. $\ln r$ for the ae-theory solution together with its value in
the corresponding Schwarzschild solution with the same  mass. In the
latter case $K=48/r^6$ in units with $M=1$, so $\log K = -6\ln r
+\ln 48$. The rate $d\ln K/d\ln r$ for the ae-theory solution seems
to alternate between roughly $-6$ and $-4.5$. The location of the
transitions may be correlated with the oscillations discussed above.
\begin{figure}
\begin{center}
 \includegraphics[angle=270,width=12cm]{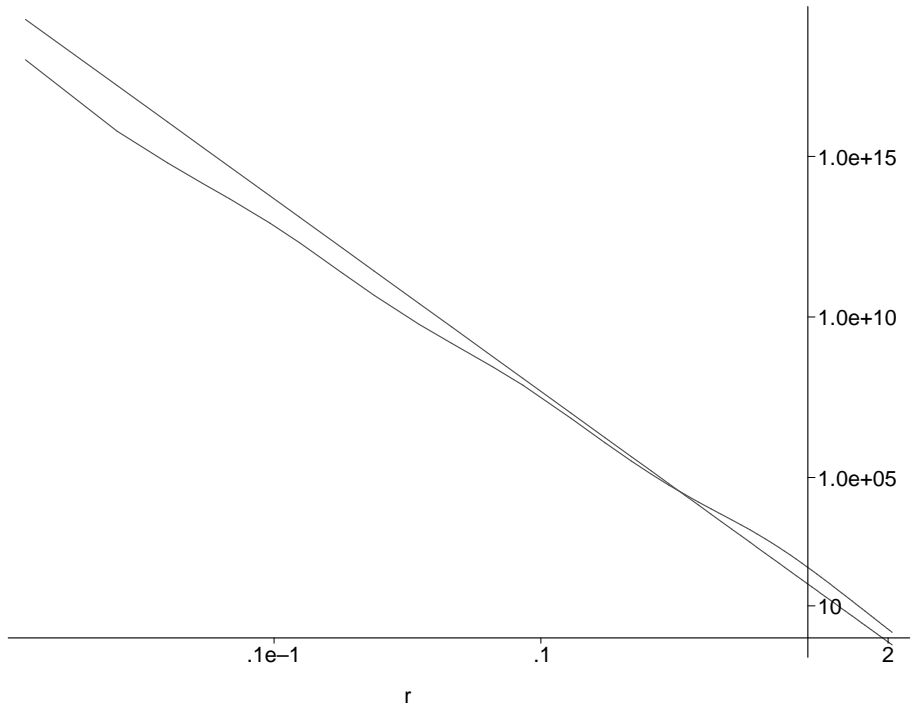}\\
\caption{\label{fig:curvature}Plot of $\ln R_{abcd} R^{abcd}$ vs.
$\ln r$ for $c_1 =0.3$ ae-theory black hole (wiggly curve)
and Schwarzschild black hole (stright line)
of the same mass, in units with $M=1$. $r$ is plotted
on a logarithmic scale. The slope for the GR case is $-6$, while
for the ae-theory case it alternates between roughly $-6$ and
$-4.5$.}
\end{center}
\end{figure}
The location of the singularity seems to be at $r=0$ for all the
values of $c_1$, although our numerical solutions do not
permit a determination of the exact
location.

\subsubsection{Aether congruence}
\label{aecong}

The aether field defines a congruence of radial timelike curves at
rest at infinity and flowing into the black hole. It is interesting
to compare this with the static frame and with the congruence of
freely falling radial geodesics with 4-velocity $v^a$ that are also
at rest at infinity. Being unit vector fields, at each point $u^a$
and $v^a$ can be fully characterized by their Killing energy, i.e.
their inner product with the Killing vector. The free-fall
congruence has a conserved energy that is equal to one if the
Killing vector is normalized to one at infinity. The aether does not
fall as quickly outside the black hole. In fact it remains rather
aligned with the Killing vector up until quite close to the horizon.

To characterize and contrast the free-fall and aether congruences we
plot in Figure \ref{vstatic} the derivative $dr/d\tau$ of radius
with respect to proper time along each congruence. Let us call this
the quantity the ``proper velocity".\footnote{The magnitude of this
quantity is affected both by the radial motion and the behavior of
the proper time. For instance as the particle becomes lightlike the
proper time goes to zero and this derivative diverges. However we
could think of no better measure of the radial velocity. One might
use the 3-velocity relative to a static observer outside the black
hole, but since the static observer becomes lightlike at the
horizon, this 3-velocity will be equal to one at the horizon for
{\it any} finite timelike 4-velocity, so it does not distinguish
different motions there.}
\begin{figure}
\begin{center}
 \includegraphics[angle=270,width=12cm]{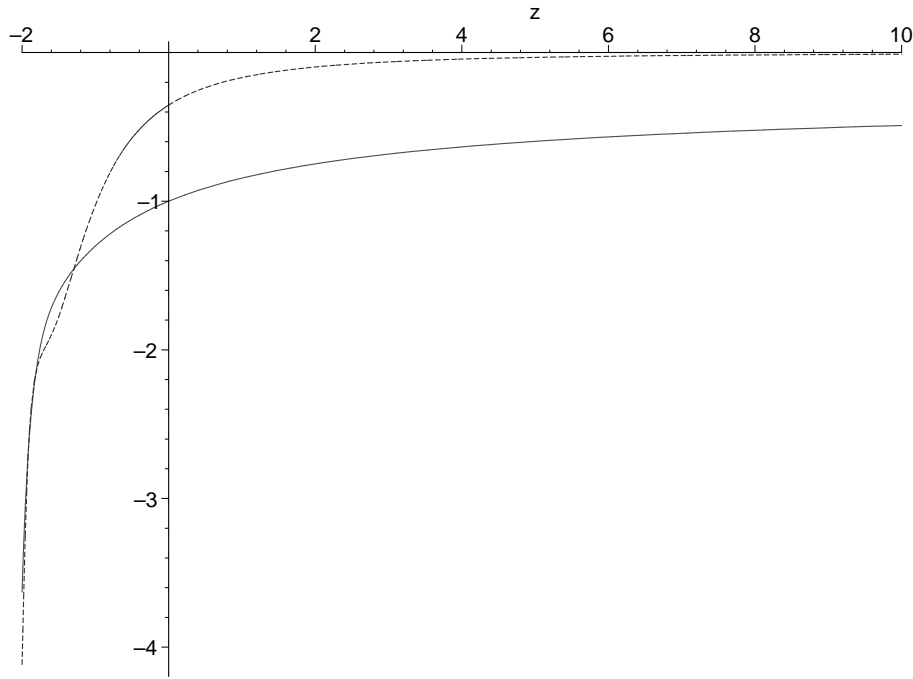}\\
\caption{\label{vstatic}Radial proper velocity $dr/d\t$ of free-fall
(lower, solid curve) and aether (upper, dashed curve) vs. $z=r-r_h$,
in units with $M=1$, for $c_1=0.3$. In contrast to the free-falling
geodesics, the aether does not begin to fall significantly inward
until close to the horizon.}
\end{center}
\end{figure}
The aether and free-fall are both at rest at infinity, but only as
the horizon is approached is the aether finally pulled away from the
Killing direction. As close as $r=3r_h$ ($z\approx4$), the proper
velocity of the aether is still about fifteen times smaller than
that of free-fall. Inside the horizon the aether proper velocity is
equal to the free-fall one around $z=-1.3$, but the 4-velocities do
not agree there. The aether is still going inward faster, but its
proper time is ``running slower" so it can have the same proper
velocity.

To compare the aether and free-fall motions inside the horizon we
plot in Figure \ref{vfreefall} the inward 3-velocity of the aether
with respect to the free-falling frame.
\begin{figure}
\begin{center}
 \includegraphics[angle=270,width=12cm]{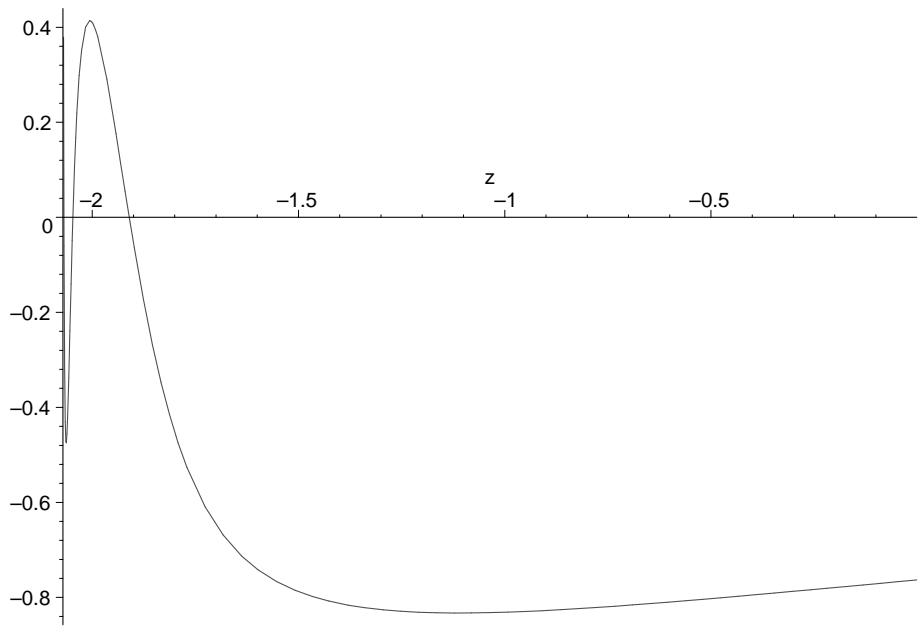}\\
\caption{\label{vfreefall}Inward 3-velocity of the aether relative
to free-fall inside the horizon for $c_1=0.3$. The velocity is
initially negative and the aether lags behind the free-fall. Near
the singularity the velocity oscillates between between faster and
slower than free-fall.}
\end{center}
\end{figure}
The relative velocity is initially negative, meaning that the aether
is not falling in as fast as the the free-fall frame. It is clear
from this plot at around $z=-1.3$ the aether still lags well behind
free-fall. However, around $z=-1.9$ the relative velocity is zero,
and after that it oscillates a couple of times (at least) before
reaching the singularity.

\subsubsection{Surface gravity and the first law of black hole
mechanics}

The laws of black hole mechanics have been shown to apply to a
wide class of generally covariant metric theories of gravity
coupled to matter~\cite{Wald:1999vt}. There appears to be no
straightforward extension of the first law and the concept of
black hole entropy to ae-theory however~\cite{Foster:2005fr}, a
difficulty that is tied to the fact that there is no smooth
extension of the aether to the bifurcation surface of the Killing
horizon. Moreover, it is not clear to which horizon the law should
apply, in a theory with multiple characteristic surfaces. For
example, in the solutions considered in this section, the spin-2
horizon is inside the spin-1 horizon which is inside the joint
spin-0 and metric horizon. One might imagine that the relevant
horizon is always the Killing horizon, but recall that by a field
redefinition we can make any one of these horizons be the Killing
horizon. We shall not try to shed any light on these puzzling
issues here. Rather, we just briefly examine the variational
relation between mass, surface gravity and area of the spin-0
horizon, for possible future use.

The first law of black hole mechanics for spherically symmetric
neutral black holes in GR takes the form
\beq \d M = \a\frac{\k \d A}{8\pi G}, \label{firstlaw} \eeq
where $A = 4\pi r_h^2$ is the horizon area, $\kappa$ is the surface
gravity, and $\a=1$. By dimensional analysis such a variational
relation must also hold in ae-theory, with some value for the
dimensionless constant $\a$ that depends on the dimensionless
coupling coefficients $c_i$. Presumably for $M$ we should put the
total energy $E$ of the spacetime, and for $``G"$ we should put the
Newton constant $G_N$ governing the attractive force between distant
bodies. Alternatively one might use the ADM mass $M_{ADM}$ and the
constant $G$ appearing in the ae-theory action (\ref{action}). As
discussed in section \ref{sec:mass}, $GM_{ADM}=G_NE=r_0/2$, so these
two choices actually yield identical ``first laws". If we express
the mass and area in terms of $r_0$ and $r_h$ respectively,
(\ref{firstlaw}) thus becomes
\beq \d r_0 =2\a\k r_h\d r_h. \label{firstlaw2} \eeq
As pointed out in section \ref{sec:mass}, $r_0$ and $r_h$ are
proportional, so we infer that $\a$ is determined by the
dimensionless combination
\beq \a = \frac{r_0}{2\k r_h^2}, \label{alpha} \eeq
which depends on the coefficients $c_i$ defining the theory.

\subsubsection{Black hole properties for different values of $c_1$}

Various properties of the black hole solutions for different values
of $c_1$ are displayed in Table \ref{bhproperties}. The other
coupling coefficients have the values $c_3=c_4=0$ and $c_2$ is given
by (\ref{c2}). For each $c_1$ there is a one-parameter family of
black hole solutions with regular spin-0 horizon, labeled by
horizon radius. For the values in the table we compare black holes
with the same horizon radius, and adopt units with $r_h=1$. The
Killing vector which enters the definition of $\k$ and $\a$ is
normalized to unity at spatial infinity.
\small
\begin{table}
\caption{\label{bhproperties} Properties of black hole solutions for
several $c_1$ values, in units with $r_h=1$.}
\begin{indented}
\item[]\begin{tabular}{@{}llllll}
\br
$c_1$&$f'(r_h)$&$\gamma = u^a v_a$&~~$r_0$&~~$\k$&~~$\a$\\
\mr
  0.1 & 2.096 & 1.619  & 0.990 & 0.507& 0.976\\
  0.2 & 2.072 & 1.608  & 0.979 & 0.517& 0.947 \\
  0.3 & 2.039 & 1.592 & 0.966 & 0.528& 0.914\\
  0.4 & 1.997 & 1.568  & 0.951& 0.543& 0.876\\
  0.5 & 1.941 & 1.535  & 0.933 & 0.562& 0.830\\
  0.6 & 1.867& 1.490  & 0.911 & 0.588& 0.787\\
  0.7 & 1.767 & 1.429 & 0.881& 0.625 & 0.704\\
\br
\end{tabular}
\end{indented}
\end{table}
\normalsize

The values of $f'(r_h)$ that yield asymptotically flat solutions for
different choices of $c_1$ are displayed in the 2nd column. These
values decrease as $c_1$ grows. For $c_1=0.8$ and larger we could
not find a $f'(r_h)$ that yielded an asymptotically flat solution.
The third column shows the gamma factor between the aether and
free-fall velocity at the horizon. The fourth column shows
$r_0=2GM_{ADM}$. This is equal to $r_h$ for $c_1=0$ (a Schwarzschild
black hole), and decreases by 12\% as $c_i$ increases up to 0.7.
Conversely, for a given mass the black hole horizon is larger for
larger $c_1$. The fifth column shows the surface gravity, which for
$c_1=0$ is $1/2r_h$ and increases by 25\% as $c_1$ increases up to
$0.7$. The last column is the dimensionless ratio (\ref{alpha})
appearing in the first law (\ref{firstlaw}), which is unity for
$c_1=0$ and decreases by 30\% as $c_1$ increases to 0.7.

\section{Discussion}
\label{discussion}

In this paper we considered the meaning of a black hole in
Einstein-Aether theory, arguing that the fastest wave mode must be
trapped if the configuration is to qualify as a causal black hole.
Regularity at the spin-0 horizon was identified as a key property of
black holes in ae-theory. It was found that, for generic values of
the coupling constants $c_i$, regularity at a metric horizon imposes
no restrictions on spherically symmetric, static local solutions but
regularity at a spin-0 horizon imposes one condition. At least for a
class of coupling constants, there is a one parameter family of
asymptotically flat black hole solutions with all horizons (metric
and spin-0,1,2) regular.

{}From an astrophysical point of view an essential question is what
happens when matter collapses. It is a plausible conjecture that
nonsingular spherically symmetric initial data will evolve to one of
the regular black holes whose existence has been demonstrated here,
but this has certainly not been shown. It would be very interesting
to answer this question by numerical evolution of the time-dependent
field equations. To do so, one could add scalar matter to form a
collapsing pulse, but this is likely not necessary since the aether
itself has a spherically symmetric radial tilt mode that can serve
the purpose. This would correspond to formation of a black hole by
an imploding spherical aether wave.

We examined several properties of regular black hole solutions for a
special class of coupling coefficients defining the ae-theory. A
complete classification of the solutions for different coefficients
remains an open research problem, as does the study of non-black
hole solutions and negative mass solutions, which strangely enough
could include black holes. Oscillating behavior approaching the
internal singularity has been identified, but not studied in detail.
Finally, only the static, spherically symmetric case was examined;
the question of rotating black hole solutions remains untouched.

\ack

We are grateful to B.Z. Foster for helpful suggestions on the
presentation. This research was supported in part by the NSF under
grant PHY-0300710.

\end{document}